\documentclass[12pt]{spieman}  
\usepackage{amsmath,amsfonts,amssymb}
\usepackage{graphicx}
\usepackage{setspace}
\usepackage{tocloft}
\usepackage{lineno}
\usepackage{dcolumn}
\usepackage{bm}
\usepackage{xcolor}

\title{Robust moir\'e flatbands within a broad band-offset range}

\author[a,b,d*]{Peilong Hong}
\author[c,+]{Yi Liang}
\author[d]{Zhigang Chen}
\author[d]{Guoquan Zhang}

\affil[a]{School of Optoelectronic Science and Engineering, University of Electronic Science and Technology of China (UESTC), Chengdu, Sichuan 611731, China}
\affil[b]{School of Mathematics and Physics, Anqing Normal University, Anqing, Anhui 246133, China}
\affil[c]{Guangxi Key Lab for Relativistic Astrophysics, Center on Nanoenergy Research, School of Physical Science and Technology, Guangxi University, Nanning, Guangxi 530004, China}
\affil[d]{The MOE Key Laboratory of Weak-Light Nonlinear Photonics, School of Physics and TEDA Applied Physics Institute, Nankai University, Tianjin 300457, China}

\cftpagenumbersoff{figure}
\cftpagenumbersoff{table}

\begin{document} 
\begin{spacing}{1.2}   

\maketitle

\begin{abstract}
Photonic analogs of the moir\'e superlattices mediated by interlayer electromagnetic coupling are expected to give rise to rich phenomena such as nontrivial flatband topology.
Here, we propose and demonstrate a scheme to tune the flatbands in a bilayer moir\'e superlattice by employing the band offset.
The band offset is changed by fixing the bands of one slab while shifting those of the other slab, which is accomplished by modifying the thickness
of the latter slab.
Our results show that the band-offset tuning not only makes some flatbands emerge and disappear, but also leads to two sets of flatbands that are robustly formed even with the change of band offset over a broad range.
These robust flatbands form either at the AA-stack site or at the AB-stack site, and as a result, a single-cell superlattice can support a pair of high-quality localized modes with tunable frequencies.
Moreover, we develop a diagrammatic model to provide an intuitive insight into the formation of the robust flatbands.
Our work demonstrates a simple yet efficient way to design and control complex moir\'e flatbands, providing new opportunities to utilize photonic moir\'e superlattices for advanced light-matter interaction including lasing and nonlinear harmonic generation.
\end{abstract}


{\noindent \footnotesize\textbf{*}\linkable{plhong@uestc.edu.cn} \, \textbf{+}\linkable{liangyi@gxu.edu.cn}  }


\section{Introduction}

Moir\'e physics is a nascent yet exciting research direction that has led to important discoveries in various areas, ranging from electronics, to optics, and to acoustics~\cite{NatRevMat2021Andrei,APL2021Chen,deng2020magic,du2023Science}.
Moir\'e physics is associated with the emergence of novel phases that are not present in the individual lattices of the superlattice, leading to intriguing physical phenomena. For instance, when two monolayer materials are brought into contact, the moir\'e potentials have been predicted and demonstrated to strongly modify the optical properties of the bilayer materials~\cite{urbaszek2019News,tran2019Nature, seyler2019Nature, alexeev2019Nature, Nat2019Jin}.
Indeed, the initial breakthrough was made in condensed matter systems by discovering exotic phenomena in moir\'e superlattices, including unconventional superconductivity~\cite{Nat2018Cao}, moir\'e excitons~\cite{tran2019Nature, seyler2019Nature, alexeev2019Nature, Nat2019Jin}, anomalous Hall ferromagnetism~\cite{PRL2020Bultinck}, to name just a few. 
These novel moir\'e effects are discovered in superlattices with appropriate interlayer coupling, but the realization of the nontrivial superlattice needs fine tuning of the 2d materials, as initiated by the seminal work on twisted bilayer graphene~\cite{Nat2018Cao,Nat2018Cao2}.
The difficulty arises from the reliance of the moir\'e effects on the formation of flatbands that only occur at magic angles in the superlattices~\cite{Nat2018Cao,Nat2018Cao2,PNAS2011Bistritzer}. Later, the concept of moir\'e physics was introduced into the realm of optics, where flexible control on superlattices is feasible~\cite{APL2021Chen,du2023Science}.
The tunability of photonic superlattices is particularly beneficial for exploring flatband physics and relevant photonic applications.

In photonics, several research groups have studied intriguing moir\'e physics with mismatched photonic lattices~\cite{PRL2020Wang,PRR2022Nguyen}, and twisted bilayer photonic slabs~\cite{PRL2021Dong,Nat2020Hu,PRL2021Lou,PRB2021Oudich,LSA2021Tang}.
The moir\'e bands are typically tuned either by twisting the bilayer slabs or by modifying the optical distance between the two slabs.
Especially, the flatbands relevant to novel moir\'e physics are found at photonic magic angles~\cite{PRL2021Dong,LSA2021Tang}, demonstrating a striking similarity with the electronic twisted bilayer graphene.
Besides, a flatband can also emerge by setting the optical distance to specific values, as discovered in the 2D twisted superlattices~\cite{PRL2021Dong,PRB2021Oudich} and 1D mismatched superlattices~\cite{PRR2022Nguyen}.
The appearance of flatbands underlies various optical phenomena such as topological transition of optical dispersion contours~\cite{Nat2020Hu}, non-Anderson-type localization of light~\cite{Nat2020Wang}, twisting-induced optical solitons~\cite{NP2020Fu}, and moir\'e quasi-bound states in the continuum~\cite{PRL2022Huang}.  
Obviously, flatbands play an essential role in exploring novel physics in optics.
Besides, a flatband mode is typically localized~\cite{APX2018Leykam,NP2020Tang}, making it highly valuable for manipulating light-matter interaction such as lasing~\cite{NN2021Mao} and optical harmonic generation~\cite{OL2022Hong}.
Therefore, flatband formation lies at the heart of intriguing moir\'e physics and relevant applications.
It is desirable and critical to develop efficient strategies to tune the moir\'e flatbands.

In this work, we employ the band offset in the band domain as an efficient knob to tune the flatbands in a mismatched bilayer superlattice.
The band offset is changed by fixing the bands of one slab while shifting those of the other slab, which are realizable by modifying the thickness of the latter slab in practice.
Remarkably, the band offset not only triggers the appearance and disappearance of a few flatbands, but also leads to two sets of flatbands that can robustly form within a broad band-offset range.
By taking advantage of the robustly formed flatbands, we further demonstrate a doubly-resonant single-cell superlattice with localized modes originating from the flatbands, and the frequencies of these modes are tunable by band offset.
Such localized modes hold great promise for manipulating advanced light-matter interaction.
Moreover, we develop a diagrammatic model to provide an intuitive insight into the formation of these robust flatbands, which can inspire new design approaches for moir\'e superlattices with tailored flatbands.
Our work thus provides an efficient way to understand and control the formation of flatbands.
Since the robust flatbands can be achieved without requiring strict magic configuration, they may have great potential in relevant applications based on moir\'e devices.

\section{Results}

\subsection{Scheme and adjustable band offset}

\begin{figure}[h]
	\centering
	\includegraphics[width=0.7\textwidth]{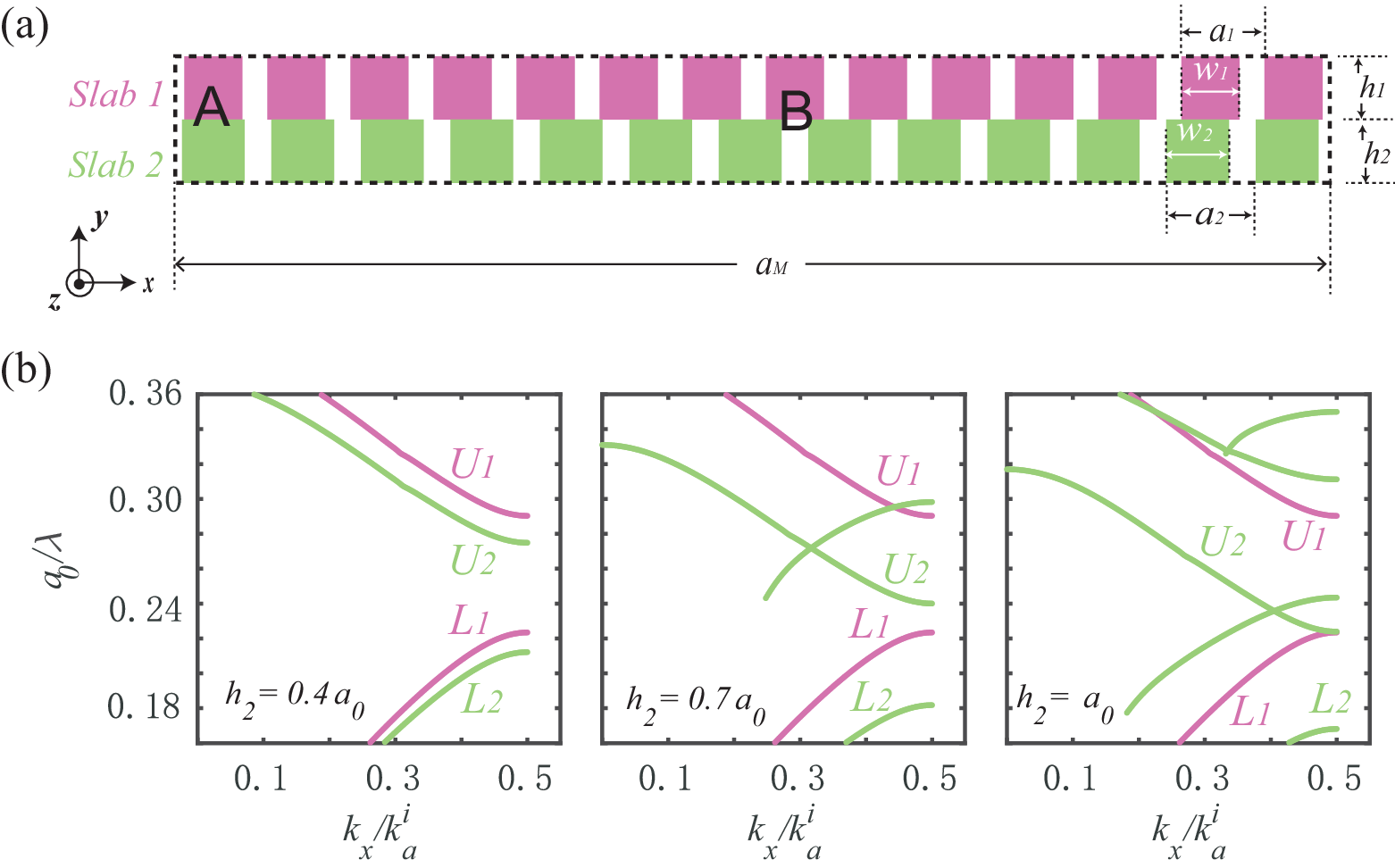}
	\caption{(a) Schematic diagram of a silicon moir\'e superlattice.
		(b) band offset adjusted by the thickness ($h_2$) of slab 2. Here, the bands are calculated only for single slab, in the absence of the other slab. $k_a^i = 2\pi/a_i (i=1,2)$.
        Note that more than two bands for slab 2 emerge within the interested region by adjusting $h_2$, but only two main bands $L_i$ and $U_i$ are marked.
		\label{Figure_1}}
\end{figure}

Our superlattice is constructed by stacking two mismatched silicon slabs in a commensurate configuration shown in Figure~\ref{Figure_1}(a). The unit size of slab 1 is $a_1 = 2N/(2N+1) a_0$, while that of slab 2 is $a_2 = 2(N+1)/(2N+1) a_0$. Here, $N$ is an integer (fixed to be 13 hereafter), and $a_0$ = 300 nm.
The resulting moir\'e superlattice has a unit size $a_M = (N+1)a_1 = Na_2$. 
Certainly, one can choose other values of $N$, and $a_M$ changes accordingly.
The width of the silicon strip is fixed to be $w_i = 0.7 a_i (i=1,2)$.
In the superlattice, slab-1's silicon strip can align with slab-2's silicon strip, creating the AA-stack site, indicated as “A” in Figure~\ref{Figure_1}(a). Slab-1' silicon strip can also align with the slab-2's air gap, creating the AB-stack site, indicated as “B” in Figure~\ref{Figure_1}(a).

To exploit the band offset as a degree of freedom to tune the moir\'e bands in the superlattice, we keep the thickness ($h_1$) of slab 1 fixed at 0.4$a_0$, but change the thickness ($h_2$) of slab 2 from 0.4$a_0$ to $a_0$. As a result, the bands of slab 1 remain unchanged, while the spectral positions of slab-2's bands are shifted. Hence, the band offset between the two slabs is modified. 
Hereafter, we focus on the TE bands, of which the electric field is polarized along $z$ axis.
The TE bands at different conditions are obtained by solving the wave equation
\begin{equation}
{\bm{\nabla}}  \times ({\bm{\nabla}}  \times {\bm{E}}_z) - \frac{\omega^2}{c_0^2} \epsilon_r {\bm{E}}_z = 0 \,.
\end{equation}
Here, $\omega$ is the frequency, $c_0$ is the velocity of light in the vacuum, and $\epsilon_r$ is the relative permittivity.
The refractive indices of the silicon and the air are set to be 3.47 and 1, respectively.
In this work, the wave equation is numerically solved through finite-element computation with Comsol Multiphysics.
Notably, some bands of the photonic slabs are above the light line $\omega = c_0\, k_x$, and therefore light can leak into the free space surrounding the slabs.
Consequently, the eigenfrequency is typically a complex value $\omega_r + i\gamma$, and the quality factor $Q = \omega_r/(2\gamma)$ describes how well an eigenmode is confined in the slabs.

Figure~\ref{Figure_1}(b) shows the TE bands of the two slabs at different $h_2$ ($=0.4a_0, 0.7a_0, a_0$).
For clarity, we only draw the eigenmodes with quality factors $Q > 50$, and therefore some bands look
incomplete in the figure.  
Within the spectral range of interest, slab 1 has two lowest bands ($L_1$ and $U_1$) that remain unchanged in the band domain.
Slab 2 also has two lowest bands ($L_2$ and $U_2$) at $h_2 = 0.4a_0$, and an initial band offset between $L_1$ and $L_2$ ($U_1$ and $U_2$) can be seen.
When $h_2$ increases to $0.7a_0$, $L_2$ and $U_2$ move downward, while $U_2$ intersects with another band.
As $h_2$ increases to $a_0$, $L_2$ and $U_2$ move downward further.
Clearly, the band offset between $L_1$ and $L_2$ ($U_1$ and $U_2$) increases with the increase of the thickness $h_2$.
Thus, the band offset is adjusted efficiently by scanning the thickness $h_2$ of slab 2.

\subsection{Robust flatbands}

\begin{figure}[h]
	\centering
	\includegraphics[width=0.9\textwidth]{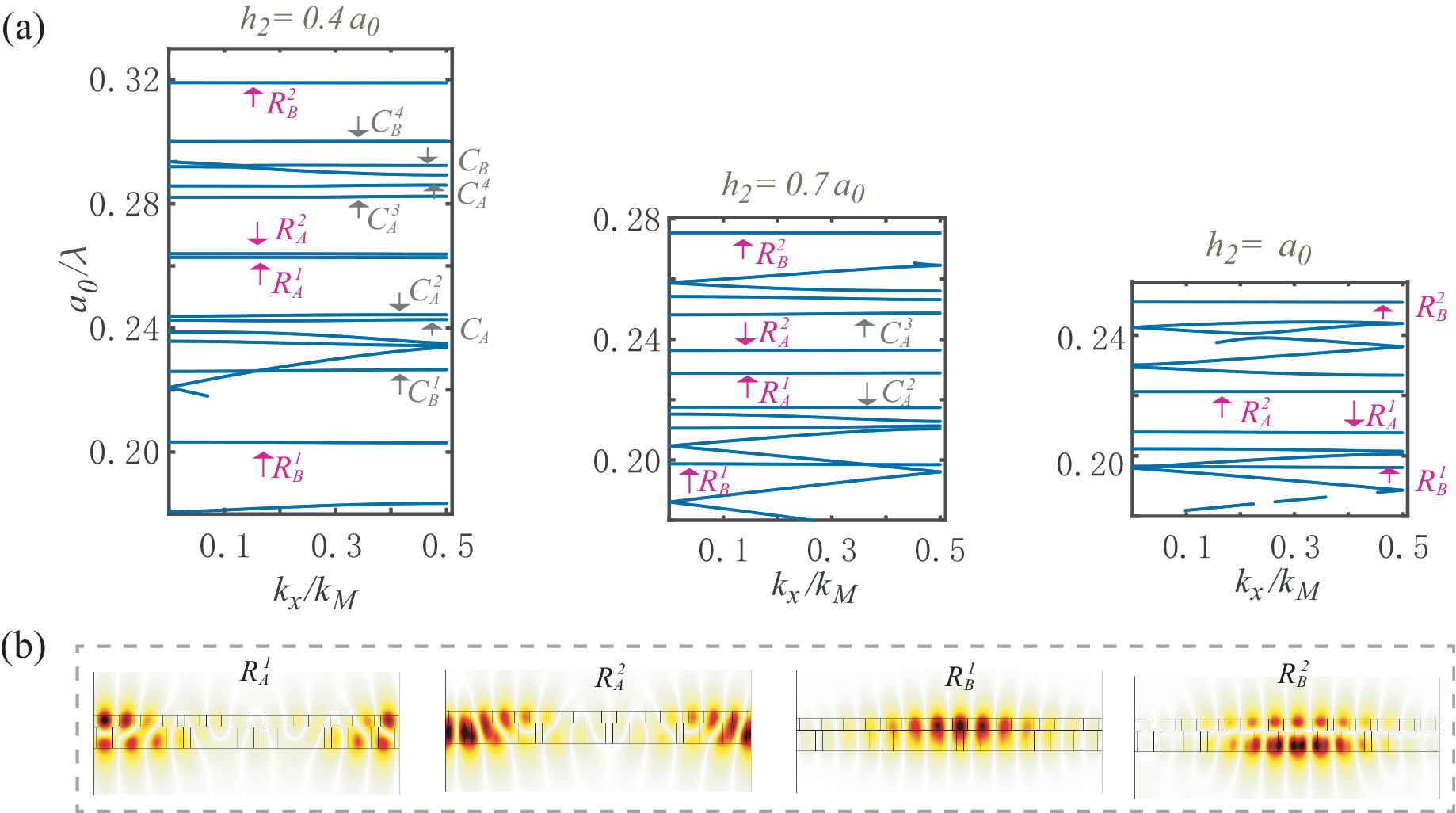}
	\caption{(a) Moir\'e bands at different band offsets. The flatbands are marked by different symbols, where $C$ labels the conventional flabands that come and go by band-offset tuning, $R$ labels the robust flatbands that preserve. The subscript A (or B) such as in $R_A^1$ denotes the center of field pattern at A (or B) site, and the superscript $i=1,2,\cdots$ denotes different flatbands.
		Here, $k_M = 2\pi/a_M$, and a same symbol at different $h_2$ indicates that the eigenmodes have similar field patterns.
		(b) The field patterns $|E_{eigen}(k_x=0)|$ of the four robust flatbands $R_A^1, R_A^2, R_B^1$ and $R_B^2$ in a single cell of a periodic superlattice at $h_2=0.7a_0$. The field magnitude is represented by a reversed hot colormap with the maximum in black and the minimum in white.
		\label{Figure_2}}
\end{figure}

Next, we investigate how the moir\'e bands are tuned by the band offset.
The moir\'e bands of the superlattice (with $h_2 = 0.4a_0, 0.7a_0 $ and $a_0$) are shown in Figure~\ref{Figure_2}(a).
The cell length $a_M$ of the superlattice is much larger, i.e. $N$ times of that of slab 2. As a result, while a band of slab 2 can extend over $2\pi/a_2$ in the $k$-space, the superlattice has mini-bands that only extend over $1/N$ of $2\pi/a_2$ in the k-space.
We have identified the flatbands that have a frequency deviation meeting the condition $(f_{max}-f_{min})/(f_{max}+f_{min})<0.15\%$.  
For clarity, the flatbands with similar field patterns are marked by a same symbol.
The results show that multiple flatbands emerge at each band offset, but some flatbands may disappear when the band offset changes. Specifically, the flatbands $C_A^1, C_A^4, C_B^1, C_B^3$ and $C_B^4$ emerge at $h_2 = 0.4 a_0$, but disappear at $h_2 = 0.7a_0$ and $a_0$. The flatband $C_A^2$ has a frequency deviation of 0.08\% at $h_2 = 0.4a_0$ and a smaller frequency deviation of 0.02\% at $h_2 = 0.7a_0$, but disappears at $h_2 = a_0$. The flatband $C_A^3$ has a frequency deviation of 0.05\% and a larger frequency deviation of 0.12\% at $h_2 = 0.7a_0$, but disappears at $h_2 = a_0$.
Such emerging and disappearing flatbands demonstrate the important role of band offset in tuning the flatbands.

Remarkably, there are four flatbands that can form robustly at different band offsets through the entire scan range, labeled as $R_A^1, R_A^2, R_B^1$ and $R_B^2$ in Figure~\ref{Figure_2}(a) for clarity.
The field patterns of these robust flatbands are shown in Figure~\ref{Figure_2}(b).
It is seen that the $R_A^1$ and $R_A^2$ modes are strongly localized around A site of the superlattice, while the $R_B^1$ and $R_B^2$ modes are strongly localized around B site of the superlattice.
The stable formation of the flatbands is quite desirable in practice, because they can be achieved without the need for a subtle magic configuration.

\subsection{ Doubly-resonant single-cell superlattice}

\begin{figure}[h]
	\centering
	\includegraphics[width=0.75\textwidth]{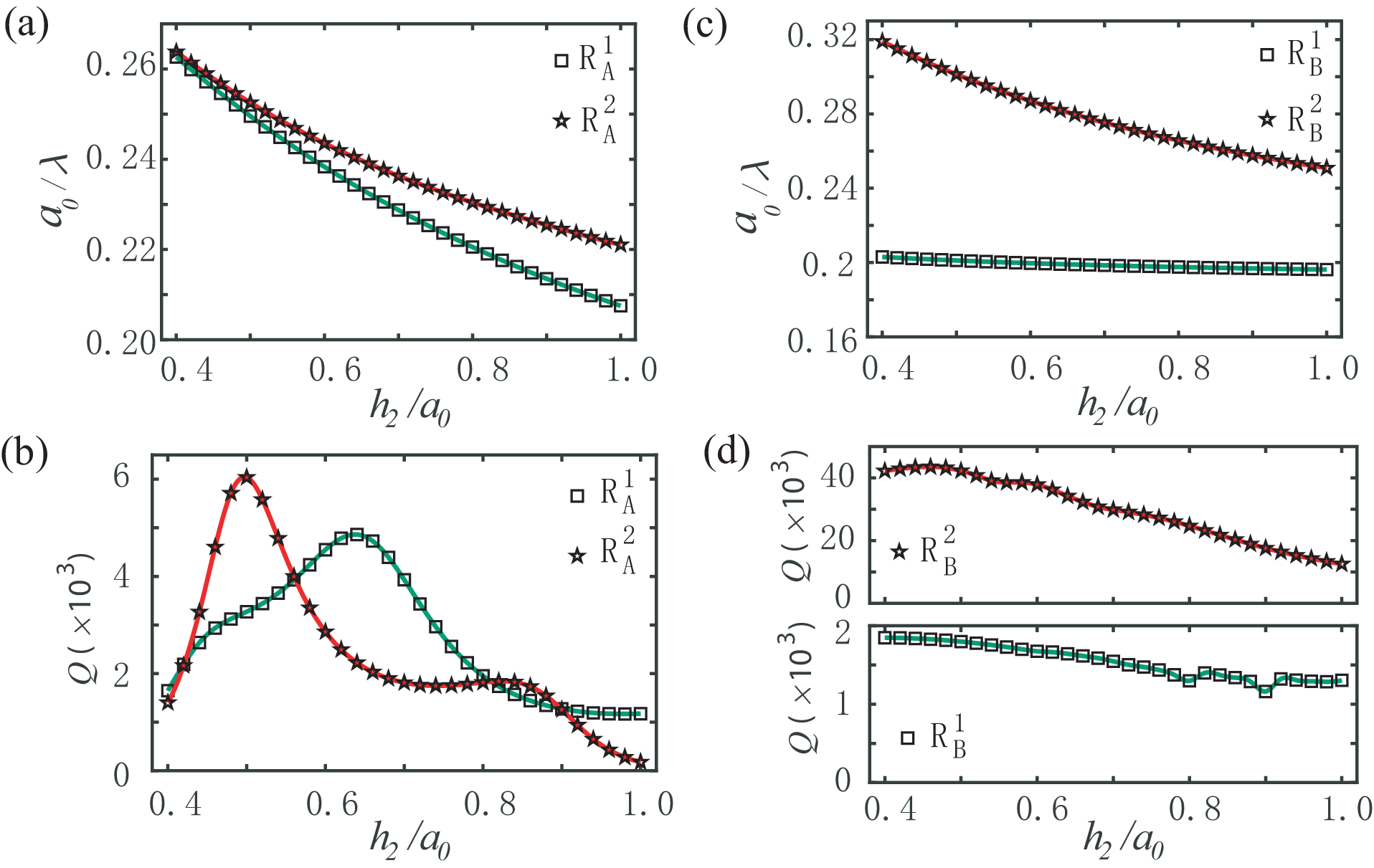}
	\caption{Tunable spectral positions of the $R_A^1$ and $R_A^2$ modes (a) and that of the $R_B^1$ and $R_B^2$ modes (c).
		The quality factors of the $R_A^1$ and $R_A^2$ modes (b) and that of the $R_B^1$ and $R_B^2$ modes (d) hold high values.
		\label{Figure_3}}
\end{figure}

The simultaneous emergence of multiple robust flatbands paves the way toward multiply-resonant superlattice with tunable frequencies, since scanning the band offset does not make these flatbands disappear but rather allows us to control their spectral positions. Particularly, these robust flatbands can even be supported by only a single-cell superlattice, due to their strongly localized wavefunctions.
We thus calculate the eigenfrequencies of localized modes with a single-cell superlattice, and focus on those originating from the robust flatbands.
The $R_A^1$ and $R_A^2$ modes are studied with an A-site centered single-cell superlattice, while $R_B^1$ and $R_B^2$ modes are studied with a B-site centered single-cell superlattice.
The spectral positions of $R_A^1$ and $R_A^2$ modes at different $h_2$ are shown in Figure~\ref{Figure_3}(a), and those for $R_B^1$ and $R_B^2$ modes are shown in Figure~\ref{Figure_3}(c). The $R_A^1$ and $R_A^2$ modes are nearly degenerate at 0.4$a_0$, and gradually separate when $h_2$ increases.
In contrast, $R_B^1$ and $R_B^2$ modes start at a large spectral separation, and become closer in spectrum when $h_2$ increases.
Notably, the frequency of $R_B^1$ mode keeps almost invariant against the change of $h_2$.

Besides, we calculated the quality factor of these flatband modes at different $h_2$.
The quality factor of $R_A^1$ mode is peaked at $h_2 =0.64a_0$ with a value $\sim5000$, but keeps larger than 1000 within the entire range as shown in Figure~\ref{Figure_3}(b).
The quality factor of $R_A^2$ mode is peaked near $h_2 = 0.5a_0$ with a value $\sim6000$, but keeps larger than 1000 up to $h_2 =0.9a_0$.
For $R_B^1$ and $R_B^2$ modes shown in Figure~\ref{Figure_3}(d), the quality factor of $R_B^1$ mode decreases slightly from $\sim$2000 to $\sim$1300 within the scan range, while the quality factor of $R_B^2$ mode keeps a high value that is between $\sim$12000 and $\sim$43000.
These results show that the robust-flatband resonances hold high quality even in a single-cell superlattice, demonstrating the possibility of constructing a high-quality multiply-resonant superlattice.

\subsection{A diagrammatic model}

\begin{figure}[h]
	\centering
	\includegraphics[width=0.85\textwidth]{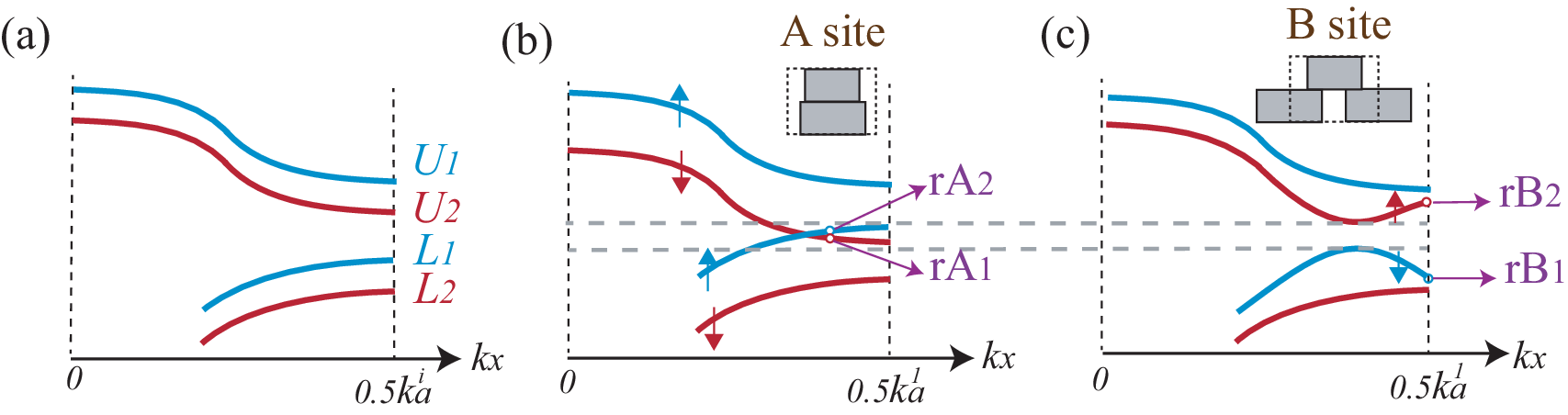}
	\caption{Two-site coupled-band diagram. (a) The band map represents the lowest two bands $L_i$ and $U_i$  ($i =1, 2$) of individual slabs of the moir\'e superlattice.
		(b) The band structure for an A-site like fictitious lattice.
		The interlayer coupling between $L_1$ and $L_2$ ($U_1$ and $U_2$) is maximized (as indicated by the arrows).
		(c) The band map of a B-site like fictitious lattice. The interlayer coupling between $L_1$ and $U_2$ is maximized.
		\label{Figure_4}}
\end{figure}

Next, we provide an intuitive insight into the formation of the robust flatbands by employing a simplified two-site coupled-band diagram.
Our model originates from the realization that the local dielectric structure continuously changes from A site to B site in the superlattice, i.e. other local positions in the superlattice have effective dielectric structures belonging to the intermediate states between A site and B site. Consequently, we consider a two-site diagram that only consists of two sets of bands related to A site and B site.
Specifically, one set of bands corresponds to a fictitious lattice with an A-site-like unit cell, and the other set of bands corresponds to a fictitious lattice with B-site-like unit cell.
The bands of the fictitious lattices originate from the interlayer coupling between the two slabs.
Under a two-mode coupling approximation, the eigenfrequencies of coupled eigenmodes are given by~\cite{Science2019Miri}
\begin{equation}
\omega^c_{\pm} = \frac{\omega_1+\omega_2}{2} \pm \sqrt{\mu^2 +( \frac{\omega_1-\omega_2}{2})^2} \,.
\end{equation}
Here $\omega_i(i=1,2)$ denotes the frequency of the eigenmode of slab $i$ such as $L_i$/$U_i$ schematically shown in Figure~\ref{Figure_4}(a), and $\omega^c_{\pm}$ denotes the frequencies of the coupled eigenmodes.
Generally, a strong interlayer coupling coefficient $\mu$ leads to a larger spectral gap between the pair of coupled eigenmodes compared to $|\omega_1-\omega_2|$.
The interlayer coupling coefficient $\mu$ is determined by the overlap of the field patterns that correspond to the pair of eigenmodes $\omega_i(i=1,2)$~\cite{IEEE1991Haus}.
Specifically, $L_i (i =1,2)$ has a spot-like field pattern with its maxima at the silicon strip, while $U_i (i =1,2)$ has a spot-like field pattern with its maxima at the air gap.
At the A site, the field maxima of $L_1$ ($U_1$) align with that of $L_2$ ($U_2$), such that their coupling coefficient $\mu$ is large.
The strong coupling increases the spectral gap between $L_1$ and $L_2$ ($U_1$ and $U_2$),
forming a coupled band diagram schematically shown in Figure~\ref{Figure_4}(b).
At the B site, the field maxima of $L_1$ align with that of $U_2$, and thus their coupling coefficient $\mu$ is large.
The strong coupling increases the spectral gap between $L_1$ and $U_2$, leading to a bandgap structure schematically shown in Figure~\ref{Figure_4}(c).
The two sets of coupled bands in Figure~\ref{Figure_4}(b) and~\ref{Figure_4}(c) constitute our two-site coupled-band diagram.

In this simplified two-site band diagram, two bands ($rA_1$ and $rA_2$) at A site always locate within the bandgap at B site. This unique band structure gives rise to optical modes that are hosted at A site but do not extend to B site. Consequently, a pair of robust flatbands always form at A site, as long as the coupled-band gap structure does not change.
This explains the robust formation of the flatbands $R_A^1$ and $R_A^2$ at A site in the superlattice.
On the other hand, the pair of coupled bands at B site ($rB_1$ and $rB_2$) reach extrema at the band edges, which are related to the robust flatbands $R_B^1$ and $R_B^2$.
Although the coupled-band edges do not locate within a bandgap at A site, they are in resonance with only one band point at A site. Moreover, the strong band coupling makes the spatial patterns of coupled eigenmodes at B site poorly overlap with those of the resonant eigenmodes at A site. As a result, the leakage to A site is weak, leading to the formation of $R_B^1$ and $R_B^2$ localized at B site. The flatbands $R_B^1$ and $R_B^2$ are stable as long as the coupled-band structure persists.
Thus, the simple two-site band diagram gives us an intuitive understanding of the formation of the robust flatbands.

\begin{figure}[h]
	\centering
	\includegraphics[width=0.85\textwidth]{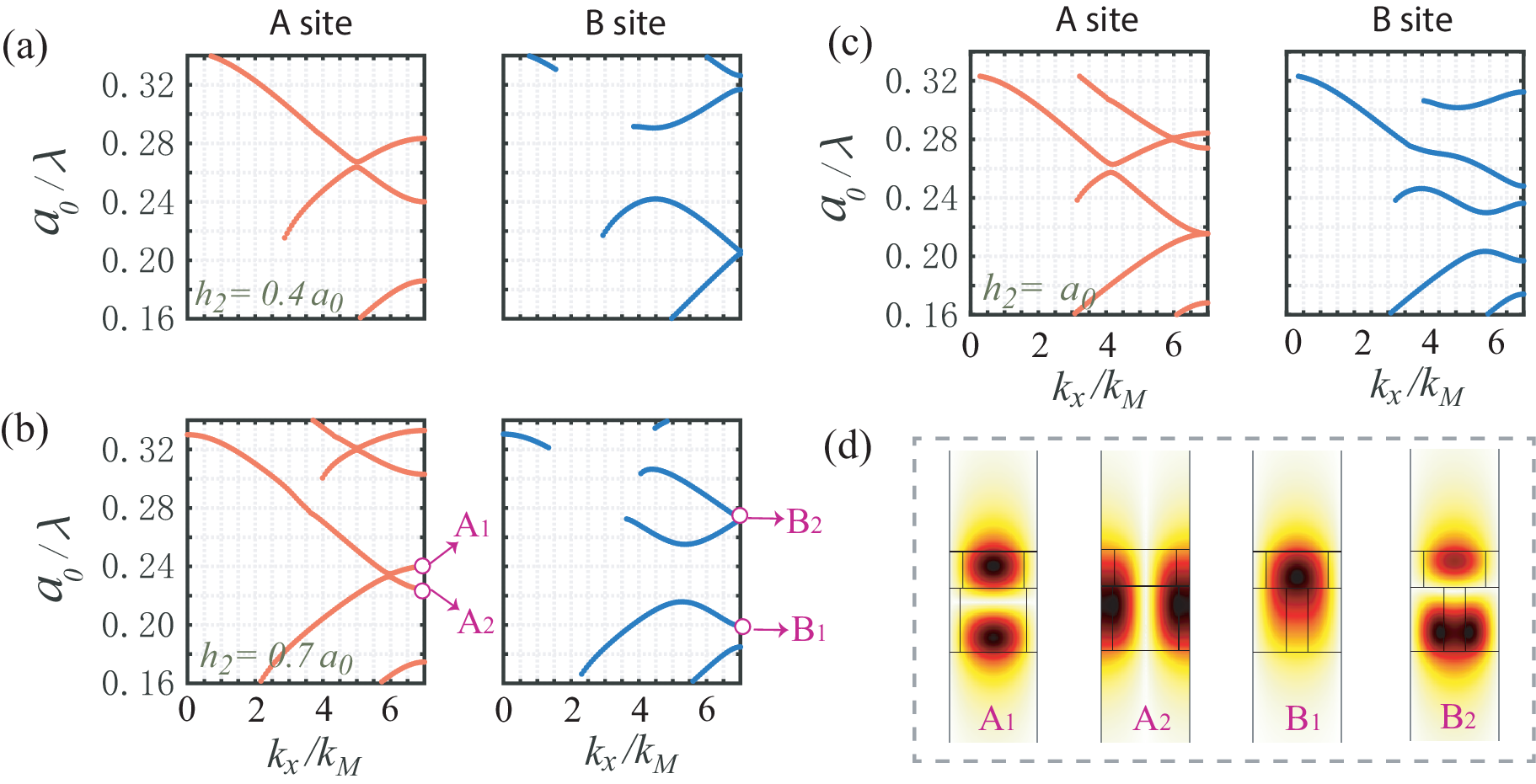}
	\caption{Coupled bands for the A-site like and B-site like fictitious lattices at $h_2 = 0.4a_0$ (a), $0.7a_0$ (b), and $a_0$ (c). The field patterns (d) for the four different band-edge modes as indicated by $A_1$, $A_2$, $B_1$ and $B_2$ in (b). Again, the field magnitude is represented by a reversed hot colormap.
		\label{Figure_5}}
\end{figure}

To further confirm the diagrammatic model, we implement full-wave calculation to obtain the accurate coupled bands of the A-site and B-site like fictitious lattices, separately. Figures~\ref{Figure_5}(a)-\ref{Figure_5}(c) show the two-site coupled bands at different $h_2$ (= $0.4a_0, 0.7a_0$ and $a_0$).
The crossing bands at A site, as well as the band gap at B site, are clearly seen in the results.
Moreover, we draw the field patterns of the four modes marked by $A_1$, $A_2$, $B_1$ and $B_2$ in Figure~\ref{Figure_5}(b), as seen in Figure~\ref{Figure_5}(d).
These field patterns are in a good agreement with the field patterns of the robust flatbands at A site ($R_A^1$ and $R_A^2$) and B site ($R_B^1$ and $R_B^2$) shown in Figure~\ref{Figure_2}(b).
This agreement adequately confirms our theoretical explanation on the formation of the robust flatbands.

\section{Discussion}

The robust flatbands differ significantly from those that are sensitive to the change of band offset, indicating the complex formation process of moir\'e flatbands.
The proposed diagrammatic model provides an intuitive understanding on the formation of robust flatbands with regard to the band offset.
Notably, the situation in twisted superlattices is much more complicated than the global band offset simply used in our work. 
When the twist angle is adjusted, locally variable band offsets are introduced in the parameter space, together with twist-dependent dielectric structures in the real space.
Nonetheless, our results may provide a straightforward approach for comprehending the formation of relatively stable flatbands in twisted superlattices~\cite{LSA2022Wang,yi2022strong}.

Besides, the diagrammatic model can guide new designs of moir\'e superlattices.
For instance, one can intentionally design two unit cells, and let them form a two-site band diagram similar to that shown in Figures~\ref{Figure_4}(b) and 4(c).
The two unit cells can be bilayer ones, as well as multi-layer ones or even monolayer ones. 
Then, a moir\'e superlattice can be constructed by connecting the two unit cells with intermediate-state cells, of which the formation of robust flatbands can be predicted under the two-site band diagram.
The field patterns of the robust flatbands can also be estimated by referring to the eigenmodes of the pair of unit cells.
Clearly, our two-site diagrammatic model significantly simplifies the design of moir\'e superlattices with tailored flatbands.

Furthermore, due to the robust formation of these flatband modes, it is possible to achieve dynamic control of their frequencies in practice.
For instance, one can fabricate a lattice with multiple domains, and each domain has a different thickness.
By shifting the multiple-domain lattice with regard to another finite-size lattice of fixed thickness, the flatband modes stably form in the bilayer region, but alter their frequencies.
Therefore, the band-offset tuning has great potential for enhancing functionalities of moir\'e photonic devices.
Certainly, structural perturbations in the superlattice could affect the flatbands, with the extent of this influence dependent on the specific fabrication conditions.
Further studies are necessary to explore and understand these effects, which may be useful for experimental implementations, especially as several experimental techniques have demonstrated their capability of realizing nontrivial moir\'e superlattices, including nanofabrication~\cite{NN2021Mao}, photorefractive effect~\cite{NP2020Fu,Nat2020Wang}, and femtosecond-laser writing~\cite{arkhipova2023PRL}.


\section{Conclusion}
In conclusion, we have demonstrated that the band offset can be an efficient knob to tune the flatbands in a moir\'e superlattice.
The band offset not only makes a few flatbands emerge and disappear,
but also leads to multiple robust flatbands with their wavefunctions localized at different stacking sites.
These robust flatbands offer promising avenues for constructing multiply-resonant moir\'e superlattices with tunable frequencies, as illustrated by the analysis of a single-cell superlattice.
Moreover, we have developed a diagrammatic model that provides an intuitive insight into the formation mechanism of the two sets of robust flatbands,
which can inspire new designs of moir\'e superlattices.
Our scheme may be further developed with AI-empowered techniques, which may prove relevant to next-generation device designs for nanophotonics~\cite{chen2021eLight}.
In particular, this work represents an important step toward controlling and understanding complex flatbands in moir\'e superlattices, and may bring about new opportunities for exploiting moir\'e superlattices in manipulating advanced light-matter interactions such as lasing~\cite{NN2021Mao}, nonlinear harmonic generation~\cite{OL2022Hong}, and enhanced free-electron radiation~\cite{yang2023photonic}.


\subsection*{Disclosures}
The authors declare no conflict of interest.

\subsection* {Acknowledgments}
This work was supported by Sichuan Science and Technology Program (2023NSFSC0460), the Fundamental Research Funds for the Central Universities (ZYGX2020J010), the Open Project Funding of the Ministry of Education Key Laboratory of Weak-Light Nonlinear Photonics (OS22-1), Guangxi Natural Science Foundation (2020GXNSFAA297041), the National Key R\&D Program of China (2022YFA1404800), and the National Natural Science Foundation (12134006).

\subsection* {Data Availability} 
The data that support the findings of this article are not publicly available. They can be requested from the corresponding authors upon reasonable request.


\bibliography{literature.bib}   
\bibliographystyle{spiejour}   



\vspace{1ex}
\noindent Biographies and photographs of the authors are not available now.


\end{spacing}
\end{document}